 \documentclass[11pt,a4paper]{article}

 \usepackage{amsmath,amssymb,graphicx,subfig,jheppub}

 \def\eq#1{Eq.~(\ref{#1})}
 
 \def\beq{\begin{equation}}
 \def\eeq{\end{equation}}
 \def\beqa{\begin{eqnarray}}
 \def\eeqa{\end{eqnarray}}
 \newcommand{\e}{\epsilon}
 \def\one{\!\!{\hbox{ 1\kern-.8mm l}}}
 \newcommand{\bra}[1]{\langle{#1}|}
 \newcommand{\ket}[1]{|{#1}\rangle}
 \newcommand{\braket}[2]{\langle{#1}|{#2}\rangle}
 \newcommand{\ex}[1]{{\rm e}^{#1}}
 \def\ii{{\rm i}}

\title{Multi-loop open string amplitudes and their field theory limit.}

\author[a]{Lorenzo Magnea}
\author[b]{Sam Playle}
\author[c]{Rodolfo Russo}
\author[d]{Stefano Sciuto}

\affiliation[a,d]{Dipartimento di Fisica, Universit\`a di Torino  \\
 and INFN, Sezione di Torino\\
 Via P. Giuria 1, I-10125 Torino, Italy}

\vspace{3mm}

\affiliation[b,c]{Centre for Research in String Theory \\ School of Physics and Astronomy\\
 Queen Mary University of London\\
 Mile End Road, London, E1 4NS,
 United Kingdom}

\vspace{5mm}

\emailAdd{lorenzo.magnea@unito.it}
\emailAdd{s.r.playle@qmul.ac.uk}
\emailAdd{r.russo@qmul.ac.uk}
\emailAdd{sciuto@to.infn.it}

 \abstract{We study the field theory limit of multi-loop (super)string amplitudes,
 with the aim of clarifying their relationship to Feynman diagrams
 describing the dynamics of the massless states. We propose an explicit
 map between string moduli around degeneration points and Schwinger
 proper-times characterizing individual Feynman diagram topologies.
 This makes it possible to identify the contribution of each light string
 state within the full string amplitude and to extract the field theory Feynman
 rules selected by (covariantly quantized) string theory. The connection between
 string and field theory amplitudes also provides a concrete tool to clarify
 ambiguities related to total derivatives over moduli space: in the superstring
 case, consistency with the field theory results selects a specific prescription
 for integrating over supermoduli. In this paper, as an example, we focus on
 open strings supported by parallel D-branes, and we present two-loop examples
 drawn from bosonic and RNS string theories, highlighting the common features
 between the two setups.}

\begin{document}
\begin{flushright}
QMUL-PH-13-06\\
\vspace*{-25pt}
\end{flushright}
\maketitle
\allowdisplaybreaks

%%%%%%%%%%%%%%%%%%%%%%%%%%%%%%%%%%%%%%%%%%%%%

 \section{Introduction}
 \label{intr}

 Multi-loop string amplitudes have been a subject of intense research for more
 than four decades, since the days of dual models~\cite{Alessandrini:1971dd},
 and studies in this field span a vast literature\footnote{See, for example,
 \cite{Verlinde:1987sd} and the review~\cite{D'Hoker:1988ta}, with the references
 therein, for a discussion of research during the eighties, and~\cite{D'Hoker:2002gw,
 Witten:2012bh} for an overview of more recent developments.}. Work in this area,
 aside from the obvious applications to string theories, has always been strictly
 related to the study of Riemann surfaces, so that it has an interest also from a
 mathematical point of view. Considering, on the other hand, the implications for
 high-energy physics, it is natural to expect that multi-loop string amplitudes should
 be connected to field theory Feynman diagrams. Indeed, in a Wilsonian sense,
 string theories reduce to quantum field theories at energies much lower than
 the scale fixed by the string tension $T = 1/(2 \pi \alpha')$. As a consequence,
 studies of the low-energy limit of perturbative string theories began very early, with
 the explicit analysis of tree-level and one-loop scalar amplitudes~\cite{Scherk:1971xy},
 and the study of gauge boson amplitudes~\cite{Gervais:1972tr}. The connection
 between string amplitudes and gauge theory amplitudes was later used also as
 a practical tool for high-energy phenomenology, with string techniques being
 used to simplify the calculation of tree-level~\cite{Mangano:1987xk} and one-loop
 gauge theory~\cite{Bern:1991aq,Bern:1993mq} and gravity~\cite{Bern:1993wt}
 amplitudes. In these simple cases, it is possible to construct a one-to-one mapping
 between the integrands of gauge theory Feynman diagrams and those of string
 amplitudes~\cite{DiVecchia:1996uq,Frizzo:1999zx,Frizzo:2000ez}, in the
 `degeneration' limit where the world-sheet turns into a graph.

 One of the aims of this paper is to provide a precise generalization of this
 correspondence to multi-loop amplitudes. We will use as a laboratory the
 calculation of string effective actions in constant background gauge fields,
 which was developed at the one-loop level in~\cite{Metsaev:1987ju,
 Bachas:1992bh}, and extended to all orders, for bosonic strings,
 in~\cite{Magnea:2004ai,Russo:2007tc}. In this paper, we will also present
 the generalization of these results to superstrings, for the case of
 Neveu-Schwarz (NS) spin structures. In the presence of a constant
 background gauge field, the string partition function is related, at
 low energies, to the so-called Euler-Heisenberg effective actions
 (see~\cite{Dunne:2004nc} for a review of this topic), and we will give a
 preliminary illustration of how known two-loop results can be recovered within
 our framework. Our string setup will involve $N$ parallel D$p$-branes in
 bosonic or type-II string theories. In both cases, the bosonic massless spectrum
 of open strings stretched between the D-branes contains a $U(N)$ gauge field,
 living in $d = p+1$ dimensions, as well as $s = D - d$ adjoint scalars (where,
 as usual, $D = 26$ and $D = 10$ for bosonic strings and superstrings respectively).
 From the space-time point of view, string configurations with $g + 1$ boundaries
 and no external legs or handles capture planar contributions to the $g$-loop
 partition function. Notice that by changing the number of coordinates with Dirichlet
 boundary conditions, and the locations of the D-branes, even from this basic
 set up one can reach an interesting set of gauge theories in different dimensions,
 coupled to adjoint scalars, as well as fermions in the superstring case.

 In order to study the field theory limit of our chosen string configuration at the
 multi-loop level, we will make use of the Schottky parametrization of (super)
 Riemann surfaces, which arises naturally in the context of the operator
 formalism~\cite{AlvarezGaume:1988bg,DiVecchia:1988cy,AlvarezGaume:1988sj},
 and is especially well-suited for the mapping between string and field
 theory quantities. We will be able to derive an explicit relationship
 between string moduli around the complete degeneration points and
 Schwinger proper times characterizing individual Feynman diagram
 topologies. We will illustrate this explicitly at two-loops, but
 the same procedure can be generalized to higher perturbative
 orders. As an example, we will consider in detail the two-loop
 contribution to the open string effective action given by a
 world-sheet with three boundaries and no handles. We will explain how
 to construct a one-to-one map, at the level of integrands, between the
 various two-loop gauge theory Feynman diagrams and different terms in
 the string amplitudes. In particular, tracing the origin of the
 various factors occurring in the string partition function to the
 functional integral over specific world-sheet fields, we will be able
 to identify the contributions of individual space-time states
 propagating in different Feynman diagram topologies. This mapping will
 enable us to identify the gauge chosen by string theory, within the
 framework of covariant quantization, extending the results
 of~\cite{Gervais:1972tr} and confirming the proposal
 of~\cite{Bern:1991an}.  Furthermore, in the superstring case, the
 precise connection between string and field theory results selects a
 specific prescription for the integration of supermoduli: in
 particular, if we envisage a higher-loop surface as obtained by gluing
 together lower-loop surfaces, then one should integrate over fermionic
 moduli by keeping fixed the (bosonic) gluing
 parameters~\cite{Witten:2013cia}. This suggests that, by using our
 approach and comparing each degeneration of a string amplitude with
 the corresponding Feynman diagrams of the low energy theory, one
 should be able to fix the total derivative ambiguities\footnote{For a
 pedagogical summary of this problem, see Section 3.4.1
 in Ref.~\cite{Witten:2012bh}.} that are always present in multi-loop
 superstring amplitudes.

 The paper is structured as follows. In section~\ref{worldsheet} we
 provide a brief discussion of the Schottky parametrization for (super)
 Riemann surfaces; we also introduce a `bra-ket' notation which
 simplifies manipulations with the (super) Schottky group. In
 section~\ref{effa} we review the form of the planar bosonic string partition
 function in the presence of a constant magnetic field and give its
 generalisation to the superstring case for the NS spin structures. In
 Section~\ref{fitl} we focus on the region of moduli space where the
 string world-sheet degenerates into the graphs corresponding to gauge
 theory Feynman diagrams. We identify a set of string moduli that are
 related in a simple way to Schwinger proper times, a connection
 which leads to an explicit algorithm to extract the field theory limit
 of the string amplitude for each graph topology. We also describe how
 one can trace the contribution of individual Feynman diagrams with the
 same topology, but with different field content, within the full
 string partition function. In section~\ref{exam}, as an example, we
 focus on the non-separating degeneration and show explicitly how a
 specific Feynman diagram with a ghost loop is obtained from the full
 integrand of the string theory amplitude in the appropriate
 degeneration limit. We perform the analysis using both bosonic strings
 and superstrings, in order to illustrate analogies and differences
 between the two formalisms. The present paper presents our method and
 the general string setting that enables us to take the field theory
 limit in a controlled way, for both bosonic strings and RNS
 superstrings. We leave to forthcoming papers a detailed analysis of
 how different gauge theories can be reached within this framework,
 including a complete study of issues related to renormalization and
 gauge-fixing.

 \section{A parametrization for (super) Riemann surfaces}
 \label{worldsheet}

 Our goal in the present paper is to describe planar interactions among
 D$p$-branes, therefore we will focus on open string world-sheets with
 boundaries but no handles. In the simplest case, the relevant world-sheet
 has the topology of the disk, which can be conformally mapped to
 $\overline{ \mathbb C}^+$, the upper-half part of the complex plane plus
 the point at infinity, with the real line representing the boundary. More
 complicated Riemann surfaces can be described  by using a construction
 due to Schottky: let us briefly summarize this formalism in the case of
 planar open string world-sheets, first in the bosonic and then in the
 supersymmetric cases.

 A projective transformation maps $\overline{\mathbb C}^+$ to itself, and can
 be represented by a $PSL(2,{\mathbb R})$ matrix $S$
 \beq
   S = \left(
   \begin{array}{cc}
     a & b \\ c & d
   \end{array} \right) \, , ~~~
   S : z \rightarrow z' = \frac{a z+ b}{c z + d} \, , ~~~ \mbox{with} ~~~~
   a d - b c = 1 \, .
 \label{eq:PSL2R}
 \eeq
 Clearly, it is convenient to introduce projective coordinates $(z_u, z_d)$, with
 $z \equiv z_u/z_d$ when  $z_d \neq 0$. We will be interested in projective
 transformations with two distinct eigenvectors $(u_u, u_d)^{\rm t}$ and
 $(v_u, v_d)^{\rm t}$, called fixed points\footnote{For the sake of simplicity,
   when $z_d \not= 0$, we can choose the representative with $z_d = 1$, but
   one should keep in mind that the bra and ket introduced here are projective
   objects, which can appear only in relations that are unchanged when they are
   rescaled.}, and an eigenvalue $\sqrt{k} < 1$, where $k$ is called multiplier
 (since $\det S = 1$, the other eigenvalue must then equal $1/\sqrt{k}$). A nice
 way to describe the action of these transformations is to use the following
 bracket notation for the points of the Riemann surface
 \beq
   | z \rangle = \left(
   \begin{array}{c}
     z_u \\ z_d
   \end{array} \right) \, , ~~~~
   \langle z | \equiv \left[\left(
   \begin{array}{cc}
       0 & 1 \\ - 1 & 0
   \end{array} \right) \left(
   \begin{array}{c}
     z_u \\ z_d
   \end{array}\right)\right]^{\rm t} \, \equiv \,
   \left[ I \ket{z} \right]^{\rm t} \, = \, (z_d, - z_u) \, .
 \label{eq:braketb}
 \eeq
 With this definition of the bra-vector we can follow the notation of~\cite{Witten:2012ga},
 and introduce a skew-symmetric bilinear form $\langle w | z \rangle$ which is proportional
 to the difference between the coordinates of the two points. Indeed, $\braket{w}{z} \equiv
 z_u w_d - z_d w_u = - \braket{z}{w}$. Therefore, if $z_d, w_d \not= 0$, $\braket{w}{z} =
 z_d w_d \, (z - w)$. In this language, we can write a projective transformation $S$ in terms
 of its multiplier $k$, and of the fixed-point kets $\ket{u}$ and $\ket{v}$, as
 \beqa
   S & = & \one \, + \, \frac{1}{\langle {v}|{u} \rangle}
   \left[ \left( \sqrt{k} + 1 \right) \ket{v} \bra{u}
   - \left(\frac{1}{\sqrt{k}} + 1 \right) \ket{u} \bra{v} \right] \nonumber \\
   & = & - \, \frac{1}{\sqrt{k}} \left(\one \, + \, \frac{1- k}{\braket{v}{u}}
   \ket{v} \bra{u} \right) \, ,
 \label{eq:projb}
 \eeqa
 where the second form is obtained by using $\one = (\ket{u} \bra{v} - \ket{v}
 \bra{u})/\braket{v}{u}$. The sign of the square root of $k$ is immaterial, since
 both choices define the same projective transformation (the situation will be
 different in the supersymmetric case). It is easy to verify that $S$ turns into $S^{-1}$
 under the exchange  $\ket{u} \leftrightarrow \ket{v}$ and that the bra corresponding
 to the ket $\ket{S z} = S \ket{z}$ is simply $\bra{S z} = \bra{z} S^{-1}$, so that the
 bilinear form is invariant under projective transformations: $\braket{S z}{S w} =
 \braket{z}{w}$. A single bracket, however, is not a well-defined object, as it depends
 on the representative chosen for $z$ and $w$; as is well-known, one can form the
 first projective invariant by using four points, since in this case all $z_d$ components
 cancel in the ratio
 \beq
   (z_1, z_2, z_3, z_4) \, = \, \frac{\braket{z_1}{z_2} \braket{z_3}{z_4}}{\braket{z_3}{z_2}
   \braket{z_1}{z_4}}  \, =  \, \frac{(z_2 - z_1)(z_4 - z_3)}{(z_2 - z_3)(z_4 - z_1)} \, .
 \label{eq:invquart}
 \eeq
 The real projective transformations we have just introduced define an isometry
 between two half-circles ${\cal C}$ and ${\cal C}'$ in the complex plane, each
 centered on the real axis. For $c \neq 0$, the half-circles are centered respectively
 in $a/c$ and $- d/c$, and both have radius $1/|c|$; if $c = 0$, we can choose
 ${\cal C}$ to be centered around the fixed point $u$, with radius $\sqrt{k}$, and
 ${\cal C}'$ to be its image under $S^{-1}$. We are now in a position to describe
 a Riemann surface with $g + 1$ boundaries and no handles, by giving $g$
 projective transformations $S_\mu$, defining $2 g$ non-overlapping half-circles
 $\{{\cal C}_\mu,{\cal C}'_\mu\}$.  The $g$ projective transformations $S_\mu$
 freely generate the $g$-loop Schottky group ${\cal S}(g)$, whose elements are
 arbitrary finite products of the $S_\mu$'s and their inverses. The genus-$g$
 Riemann surface $\Sigma_g$ is then obtained by cutting away the interior of the
 disks $\{{\cal C}_\mu, {\cal C}'_\mu\}$ and by imposing the equivalence relation\footnote{The choice of the $g$ generators $S_\mu$ induces a canonical homology basis on $\Sigma_g$: going around a cycle $a_\mu$ corresponds to going around the isometric circle ${\cal C}_\mu$ or ${\cal C}'_\mu$, while moving on a path that brings from a point $z\in{\cal C}_\mu$ to the point $S_\mu(z)\in{\cal C}'_\mu$ corresponds to going around a  $b_\mu$  cycle.}
 $z \cong T(z)$, $\forall \, T\in {\cal S}(g)$.

 At this point it is easy to derive the dimension of moduli space in this representation:
 each $S_\mu$ contains three real parameters, but we are free to choose coordinates
 in $\overline{\mathbb C}^+$ by using a projective transformation $S_0$. In the new
 coordinates the Schottky generators change by a similarity transformation, as  $S_\mu'
 = S_0 S_\mu S_0^{-1}$, so we are free to `gauge away' three parameters among the $2
 g$ fixed points $u_\mu$ and $v_\mu$. As a consequence, for $g \geq 2$, the dimension
 of the bosonic moduli space is $3 g - 3$, a well-known result.

 This construction can be generalized straightforwardly to the supersymmetric case,
 as described for instance in~\cite{Martinec:1986bq,Neveu:1987mn,DiVecchia:1988jy}.
 To formulate the basic concepts in our notation, we will use boldface letters to indicate
 superspace coordinates and Greek letters to indicate Grassmann variables. To
 describe super-projective transformations, we generalize our bracket notation to
 three-dimensional vectors with one anti-commuting component, as
 \beq
   | \mathbf{z} \rangle \, = \, \left(
   \begin{array}{c}
     z_u \\ z_d \\ \hat\chi
   \end{array}
   \right) \, , ~~~~
   \langle z| \equiv \left[ \left(
   \begin{array}{ccc}
       0 & 1 & 0 \\ - 1 & 0 & 0 \\ 0 & 0 & 1
   \end{array}
   \right) \left(
   \begin{array}{c}
     z_u \\ z_d \\ \hat\chi
   \end{array}
   \right)\right]^{\rm t} \equiv \, \Big[\mathbf{I} \,
   \ket{\mathbf{z}} \Big]^{\rm t} \, = \, (z_d,-z_u,\hat\chi) \, ,
 \label{eq:braketf}
 \eeq
 where we define $\mathbf{z} = (z = z_u/z_d, \chi = \hat\chi/z_d)$, since both
 bosonic and fermionic coordinates are projective variables. The bilinear form
 is then given by $\braket{{\mathbf z}_1}{{\mathbf z}_2} \equiv z_{2 u} z_{1 d} -
 z_{2 d} z_{1 u} - \hat\chi_2 \hat\chi_1 =  - \braket{{\mathbf z}_2}{{\mathbf z}_1}$;
 thus, when $z_{1 d}, z_{2 d} \not= 0$, we can define the superspace difference
 $\mathbf{z}_2 - \mathbf{z}_1 \equiv z_2 - z_1 - \chi_2 \chi_1$, which allows us
 to write $\braket{{\mathbf z}_1}{{\mathbf z}_2} = z_{1 d} z_{2 d} (\mathbf{z}_2 -
 \mathbf{z}_1)$.  A super-projective transformation $\mathbf{S}$ can now be
 parametrized, in bracket notation, in terms of its multiplier and its two (even) fixed
 points\footnote{The third fixed point is Grassmann-odd; even if it does not enter in
   the calculations, its presence is important because it prevents the generalization
   of the identity $\one = (\ket{u} \bra{v} - \ket{v} \bra{u})/\braket{v}{u}$ from holding in
   the supersymmetric case.}, as
 \beq
   \mathbf{S} \, = \, \one \, + \, \frac{1}{\langle \mathbf{v}|\mathbf{u} \rangle}
   \left[ \left(1- \ex{\ii \pi \varsigma} \, k^{\frac{1}{2}} \right) \ket{\mathbf{v}}\bra{\mathbf{u}}
   \, - \, \left(1-\ex{\ii \pi \varsigma} k^{-\frac{1}{2}} \right) \ket{\mathbf{u}}
   \bra{\mathbf{v}} \right] \, ,
 \label{eq:SuSbraket}
 \eeq
 where we set $\mathbf{u} = (u,\theta)$ and $\mathbf{v} = (v,\phi)$. Notice that now the
 branch of the square root is important; for later convenience, we introduced the
 parameter $\varsigma$, which can take the values $0$ or $1$, and determines the spin
 structure along the $b$-cycles of the Riemann surface. In our conventions $k^{\frac{1}{2}}$ is negative and the trivial spin structure corresponds to $\varsigma=0$; the eigenvectors $|\mathbf{v} \rangle$ and $|\mathbf{u} \rangle$ belong to the eigenvalues $\ex{\ii \pi \varsigma} k^{\frac{1}{2}}$ and $\ex{\ii \pi \varsigma} \, k^{-\frac{1}{2}}$ respectively. Notice also that we can switch
 from the choice $\varsigma=0$ to $\varsigma=1$ simply by replacing $k \to \ex{2\pi\ii} k$.
 As in the bosonic case, $S$ turns into $S^{-1}$ under the exchange $\ket{\mathbf{u}}
 \leftrightarrow \ket{\mathbf{v}}$, so that the bilinear form is again invariant under
 super-projective transformations. Another novelty of the supersymmetric case
 is that it is possible to construct, beyond the obvious supersymmetric generalization
 of the cross-ratio defined in \eq{eq:invquart}, a non-trivial super-projective and scale
 invariant combination with only three points~\cite{Hornfeck:1987wt}. It is given by
 \beq
   \Theta_{\mathbf{z}_1 \mathbf{z}_2 \mathbf{z}_3} \, \equiv \,
   \frac{\hat\theta_1 \braket{\mathbf{z}_3}{\mathbf{z}_2} + \hat\theta_2
   \braket{\mathbf{z}_1}{\mathbf{z}_3} + \hat\theta_3
   \braket{\mathbf{z}_2}{\mathbf{z}_1}+ \hat\theta_1 \hat\theta_2
   \hat\theta_3}{\sqrt{\braket{\mathbf{z}_2}{\mathbf{z}_1} \braket{\mathbf{z}_3}{
     \mathbf{z}_2} \braket{\mathbf{z}_1}{\mathbf{z}_3}}}\;.
 \label{eq:cubsuinva}
 \eeq
 The construction of a super-Riemann surface can now proceed exactly as in the
 bosonic case. One introduces $g$ independent super-projective transformations
 $\mathbf{S}_\mu$, whose bosonic part defines non-overlapping isometric circles.
 These generators are then used to construct freely the super Schottky group
 $\overline{\cal S}(g)$. The elements of the group induce the equivalence relation
 $\mathbf{z} \cong \mathbf{T}(\mathbf{z}), \forall \, \mathbf{T} \in \overline{\cal S}(g)$.
 Again, one can easily derive the dimension of moduli space: each generator
 contains three bosonic and two fermionic variables (notice that the multiplier
 $k$ does not have a supersymmetric partner, and fermionic components are
 present only in the fixed points). The $g$ generators, minus the gauge freedom
 associated with an overall similarity transformation, yield thus $3 (g - 1)$ bosonic
 and $2 (g - 1)$ fermionic coordinates, for $g \geq 2$, as expected. We will now
 use the Schottky parametrization to construct expressions for string partition
 functions in a constant background gauge field, considering separately the
 bosonic string and the superstring cases.

 \section{Multiloop string effective actions}
 \label{effa}

 The operator formalism naturally yields expressions for string amplitudes written
 in terms of series over the genus-$g$ Schottky group ${\cal S}(g)$, introduced
 in the previous section.  Here we will focus on the two-loop case, $g = 2$, but it
 should be kept in mind that the explicit expressions written below naturally
 generalize to all orders in the genus expansion.

 For bosonic strings, the three independent parameters characterizing the genus-two
 Riemann surface can be identified as the two multipliers, together with
 one of the projective invariant cross-ratios of fixed points defined in \eq{eq:invquart},
 \beq
   k(S_1) \equiv k_1 \, , \qquad
   k(S_2) \equiv k_2 \, , \qquad
   \eta \equiv (u_1,u_2,v_1,v_2) \, , \quad \mbox{or} \qquad
   y \equiv (v_2,u_2,v_1,u_1) \, .
 \label{eta}
 \eeq
 In the bosonic case the two cross-ratios $\eta$ and $y$ are not independent, since
 $\eta + y = 1$: we can therefore parametrize the surface with either one of them.
 As pointed out in section~5.1.3 of Ref.~\cite{Witten:2012ga},
 in the NS case, the relation between the supersymmetric
 generalizations of $\eta$ and $y$ involves the product of two cubic
 invariants of the form given in~\eq{eq:cubsuinva} (see~\eq{etay}). From the counting
 argument of the previous section, we know that two of these cubic invariants are
 independent, and can be used as Grassmann moduli for the super-Riemann surface
 we are interested in. The choice between the two bosonic parameters will then have
 non-trivial consequences.

 We focus now on the construction of the string effective action in a background
 $U(N)$ gauge field. In order to proceed, we consider a stack of $N$ D-branes,
 supporting non-vanishing background field strengths $F_{\mu \nu}^{(A)}, (A = 1,
 \ldots, N)$ on their world-volume. In the case of constant (and mutually commuting)
 gauge fields, the calculation of the effective action can be written in terms of free
 two-dimensional conformal field theories, because such configurations affect the
 world-sheet theory only through a change in the boundary conditions of two-dimensional
 fields. For example, for the open string coordinates $x^\mu$ along the D-brane
 world-volume, one finds
 \beq
   \left[ \partial_\sigma x_\mu \, + \, {\rm i} \, F_{\mu \nu}^{(\sigma)}
   \, \partial_\tau x^\nu \right]_{\sigma = 0, \pi}  = \, 0 \quad \longrightarrow \quad
   \partial_{\bar z} x^\mu \Big|_{\sigma = 0, \pi}  = \,
   (R_\sigma)^\mu_{~\nu} \, \partial_z x^\nu \, \Big|_{\sigma = 0, \pi} \, \, .
 \label{boundicon}
 \eeq
 In \eq{boundicon}, we have set as usual $x_\mu (z, \bar{z}) = (X_\mu (z) +
 \tilde{X}_\mu (\bar{z}))/2$, while $F^{(\sigma)}_{\mu \nu}$ is the field living on
 the D-brane on which the endpoint $\sigma = 0$ or $\pi$ is attached. In the second
 step, we have introduced the standard complex coordinates $z = \ex{\tau + \ii
 \sigma}$, and the matrix $(R_\sigma)^\mu_{~\nu} = \left[(\eta - F^{(\sigma)})^{-1}
 (\eta + F^{(\sigma)}) \right]^\mu_{~\nu}$, where $\eta_{\mu \nu}$ is the Minkowski
 metric of $d$-dimensional space-time.  For simplicity, we choose $N = 2$ (thus
 restricting to a $U(2)$ gauge group) and focus on the case where the gauge field
 $F^{A = 1}$ is zero, while $F^{A = 2}$ is non-vanishing and constant in the $(x_1,
 x_2)$ plane, and zero in all other directions. Strings stretching between the two
 available D-branes are then charged under the background field ${\cal F}_{\mu
 \nu} = F^{A = 2}_{\mu \nu} - F^{A = 1}_{\mu \nu}$. On the string side, it is more
 useful to parametrize the external field by using the eigenvalues of $R_2$, so as
 to diagonalize the boundary conditions in~\eq{boundicon}. We write then
 \beq
   {\cal F}_{12} \, = \, {\rm diag} \{ B/2, - B/2\} \quad ; \quad \qquad
   \tan \pi \e \, \equiv \, 2 \pi \alpha' \, B \, .
 \label{epsi}
 \eeq
 For each open string, one of the two boundary conditions identifies the anti-holomorphic
 and the holomorphic coordinates, so we can express everything in terms of the latter.
 The other boundary condition fixes the monodromy of the string coordinates $\partial X$;
 in particular, strings stretched between D-branes with different background fields have
 a periodicity fixed by the parameter $\epsilon$. One finds
 \beq
   X^\pm (z) \, \equiv \, \frac{X^1(z) \pm \ii X^2(z)}{\sqrt{2}}  \quad \longrightarrow \quad
   \partial_z X^\pm \left( \ex{2 \pi \ii } z \right) \, = \, \ex{\pm 2 \pi \ii
   \epsilon} \, \partial_z X^\pm(z) \, .
 \label{eq:mono}
 \end{equation}
 In the superstring case, the world-sheet fermions $\psi^\pm (z)$ have the same monodromy
 as $\partial X^\pm(z)$, since they belong to the same (world-sheet) supermultiplet.

 The expression for the $g$-loop bosonic string partition function in this setup was
 studied in Ref.~\cite{Magnea:2004ai}. In the following we will provide a generalization of the
 results of~\cite{Magnea:2004ai} to the Neveu-Schwarz spin structure of the the RNS
 superstring. The partition functions in the two cases can be written in a compact form as
 \beq
   Z(\vec{\e}) \, = \, C_g (\vec{\e} \, ) \int \left[d m \right]^{\vec{\e}}_g \, , \qquad
   \mathbf{Z} (\vec{\e}) \, = \, C_g (\vec{\e} \,) \int \left[d \mathbf{m}
   \right]^{\vec{\e}}_g\;,
 \label{zeps}
 \eeq
 where $\vec{\e}$ is a vector with $g$ components defining the periodicity of $\partial X^\pm$
 along each $b$-cycle and $C_g (\vec{\e})$ is an overall normalization (independent of the
 world-sheet moduli), which can be computed using the results of Ref.~\cite{DiVecchia:1996uq},
 and is given explicitly in Ref.~\cite{Magnea:2004ai}. In the following, we will focus on the
 $g = 2$ case. There, the integrands in \eq{zeps} receive contributions from
 two different types of planar string diagrams: the first possibility is to have all three
 string boundaries on the same D-brane, while in the second case we have one boundary
 on a D-brane and the other two on the other D-brane. Of course only this second
 type of diagram receives contributions from charged string states and thus
 depends non-trivially on the background field ${\cal F}$. In particular, we will choose the
 description\footnote{We will follow the conventions of section~4.2 of~\cite{Russo:2007tc},
   in particular $\vec{\epsilon}=(\epsilon,-\epsilon)$.} where the external boundary of the
 diagram is on the magnetized D-brane, while the two internal boundaries are on the
 `neutral' D-brane, where $F = 0$ (of course the other possibilities are obtained simply
 by permuting the boundaries).

 To give an example of the typical form of the expressions for geometric objects in the
 Schottky parametrization, let us begin by considering  the measure of integration
 for the disk with all three boundaries lying on the same D-brane. In the bosonic case
 we write
 \beq
   \left[d m \right]^0_2 \,\, = \,\, \frac{d k_1 \, dk_2 \, d
     \eta}{k_1^2 \, k_2^2 \, (1 - \eta)^2} \,
   F_{\rm gh} (k_i, \eta) \, F_{\rm gl}^{(0)} (k_i, \eta) \, F_{\rm scal} (k_i, \eta) \, ,
 \label{aritwomeas}
 \eeq
 where we have labelled the various factors anticipating the role that they are going
 to play in the field theory limit, as discussed below. Since we are interested in studying the
 limit where the world-sheet degenerates into thin and long strips, we focus on the
 open string language. In this case one finds~\cite{DiVecchia:1987uf}
 \beqa
   F_{\rm gh} (k_i, \eta)  & = & (1 - k_1)^2 \, (1 - k_2)^2 \,\,
   {\prod_\alpha}' \prod_{n = 2}^\infty \left(1 - k_\alpha^n \right)^2 \, ,
   \nonumber  \\ \label{ghosts}
   F_{\rm gl}^{(0)} (k_i, \eta) & = &  \Big[ \det \left({\rm Im} \, \tau
   \right) \Big]^{- \frac{d}{2}} \, {\prod_\alpha}' \prod_{n = 1}^\infty
   \left(1 - k_\alpha^n \right)^{- d} \, , \\
   F_{\rm scal} (k_i, \eta) & = & {\prod_\alpha}' \prod_{n = 1}^\infty
   \left(1 - k_\alpha^n \right)^{- s} \, . \nonumber
 \eeqa
 The product ${\prod_\alpha}'$ is over all elements $T_\alpha \in {\cal S}(2)$ which
 are not integer powers of other elements, taken modulo cyclic permutations of their factors, and
with the identity excluded; $\tau$ is the period matrix of the Riemann surface, whose
 expression in the Schottky parametrization can be found, for instance, in
 Eq.~(A.14) of~\cite{DiVecchia:1988cy}. Notice that the determinant of $\tau$
 arises from integration over the zero modes of the $d$ bosonic string coordinates
 with Neumann boundary condition, while integration over non-zero modes involves
 also coordinates with Dirichlet boundary conditions. In the following it will be useful
 to keep separate, as we did in~\eqref{aritwomeas}, the contributions to the vacuum
 amplitudes that have a different world-sheet origin. In particular, the factor
 $F_{\rm gh}$ arises from the partition function of the world-sheet $(b,c)$ system;
 the factor $F_{\rm gl}^{(0)}$ results from the partition function of $d$ bosonic
 string coordinates with Neumann boundary conditions, including zero-mode and
 oscillator contributions; finally, the factor $F_{\rm scal}$ gives the contribution of
 $s$ string coordinates with Dirichlet boundary conditions.

 The superstring version of the same vacuum amplitude can be found in
 Ref.~\cite{DiVecchia:1988jy}, and can be written as
 \beq
   \left[d \mathbf{m} \right]^0_2 \,\, = \left[
   \frac{1}{dV_{\mathbf{v}_1 \mathbf{u}_1 \mathbf{v}_2}}
   \prod_{i = 1}^2  \frac{d k_i}{k_i^{3/2}} \, \frac{d \mathbf{u}_i \,
   d \mathbf{v}_i}{\mathbf{v}_i - \mathbf{u}_i}  \right]
   \mathbf{F}_{\rm gh} (k_i, \eta, \theta, \phi) \, \mathbf{F}_{\rm gl}^{(0)}
   (k_i, \eta, \theta, \phi) \, \mathbf{F}_{\rm scal} (k_i, \eta, \theta, \phi) \, .
 \label{aritwomeasf}
 \eeq
 The factor
 \beq
   \frac{1}{d V_{\mathbf{v}_1 \mathbf{u}_1 \mathbf{v}_2}} \, = \,
   \frac{\sqrt{(\mathbf{v}_1 - \mathbf{u}_1) (\mathbf{u}_1 -
   \mathbf{v}_2) (\mathbf{v}_2 - \mathbf{v}_1)}}{d\mathbf{v}_1
   d \mathbf{u}_1 d\mathbf{v}_2} \, d \Theta_{\mathbf{v}_1 \mathbf{u}_1 \mathbf{v}_2} \, \,
 \label{eq:gaugefix}
 \eeq
 takes into account the super-projective invariance of the integrand, which allows to fix
 three bosonic and two fermionic variables. As a consequence, the factor in square
 brackets in \eq{aritwomeasf} can also be written as
 \beq
   \frac{1}{dV_{\mathbf{v}_1 \mathbf{u}_1 \mathbf{v}_2}}
   \prod_{i = 1}^2  \frac{d k_i}{k_i^{3/2}} \, \frac{d \mathbf{u}_i \,
   d \mathbf{v}_i}{\mathbf{v}_i - \mathbf{u}_i} \, = \,
   \frac{d k_1}{k_1^{3/2}} \, \frac{d k_2}{k_2^{3/2}} \,
   \frac{d \mathbf{u}_2 \, d \Theta_{\mathbf{v}_1 \mathbf{u}_1
   \mathbf{v}_2}}{\mathbf{v}_2 - \mathbf{u}_2} \,
   \sqrt{\frac{(\mathbf{u}_1 - \mathbf{v}_2)(\mathbf{v}_2 -
   \mathbf{v}_1)}{\mathbf{v}_1 - \mathbf{u}_1}} \, .
 \label{ssc}
 \eeq
 As discussed in Ref.~\cite{Witten:2013cia}, it is important to specify the bosonic
 variables that are kept fixed when performing the Berezin integration over Grassmann
 variables. We will come back to this point in the next section; now we give the
 expressions in the Schottky parametrization for the objects entering in the NS
 vacuum energy, \eq{aritwomeasf}. One finds~\cite{DiVecchia:1988jy}
 \beqa
   \mathbf{F}_{\rm gh} (k_i, \eta)  & = &  \frac{(1 - k_1)^2 \, (1 - k_2)^2}{(1
   - \ex{\ii  \pi \varsigma_1} \, k_1^{\frac{1}{2}})^2 (1 - \ex{\ii  \pi \varsigma_2} \,
   k_2^{\frac{1}{2}})^2 } \, \,  {\prod_\alpha}' \prod_{n = 2}^\infty  \left(
   \frac{1 - k_\alpha^n}{1 - \ex{\ii \pi \, \vec{\varsigma} \cdot \vec{N}_\alpha} \,
   k_\alpha^{n - \frac{1}{2}}} \right)^{\! 2} \, ,
   \nonumber \\ \label{ghostsf}
   \mathbf{F}_{\rm gl}^{(0)} (k_i, \eta) & = &  \Big[ \det \left({\rm Im}\tau \right)
   \Big]^{- \frac{d}{2}} \, \,
   {\prod_\alpha}' \prod_{n = 1}^\infty \left(\frac{1 - \ex{\ii \pi \, \vec{\varsigma} \cdot \vec{N}_\alpha} \, k_\alpha^{n - \frac{1}{2}}}{1 - k_\alpha^n}\right)^{\!\! d} \, , \\
   \mathbf{F}_{\rm scal} (k_i, \eta) & = & {\prod_\alpha}' \prod_{n = 1}^\infty
   \left(\frac{1 - \ex{\ii \pi \, \vec{\varsigma} \cdot \vec{N}_\alpha} \,
   k_\alpha^{n - \frac{1}{2}}}{1 - k_\alpha^n}\right)^{\!\! s} \, . \nonumber
 \eeqa
At $g$ loops, the vectors $\vec{\varsigma}$ and $\vec{N}_\alpha$ have $g$-components. The vector $\vec{\varsigma}$, whose components can take the values 0 and 1, defines the spin-structure along the $b$-cycles of the  homology basis of the Riemann surface; $\vec{N}_\alpha$, on the other hand, has integer-valued entries: the $\mu$-th entry counts how many times the generator $S_\mu$ enters in the element of the Schottky group $T_\alpha$, with $S_\mu$ contributing $1$, while $S_\mu^{-1}$ contributes $-1$.

However, even for $\vec{\varsigma}=0$, there is still a sign ambiguity in the half-integer powers of $k_\alpha$. It is fixed in the following way: for $\vec{\varsigma}=0$, choose all $k_{\mu}^{1/2}$ negative for $\mu=1,2,\ldots,g$; then, always for  $\vec{\varsigma}=0$, $k_{\alpha}^{1/2}$ is defined as the smallest (in absolute value) of the three eigenvalues of $T_{\alpha}$.  For instance, for the element $\mathbf{S}_1 \mathbf{S}_2^{-1}$, we have $\big[k(\mathbf{S}_1 \mathbf{S}_2^{-1})\big]^{{1}/{2}}=k_1^{{1}/{2}} k_2^{{1}/{2}} \cdot ( \mathbf{u}_1, \mathbf{v}_1, \mathbf{v}_2,\mathbf{u}_2) + {\cal O}(k_i)$ and the sign of the harmonic ratio determines that of $[k(\mathbf{S}_1 \mathbf{S}_2^{-1})]^{{1}/{2}}$. With the choice of fixed points used later in Section~\ref{exam} ($\mathbf{u_1} = (0, 0)$, $\mathbf{v_1} = (\infty, 0)$, $\mathbf{u_2} = (u, \theta)$, and $\mathbf{v_2} = (1, \phi)$ with $0<u<1$), the harmonic ratio appearing above is negative; therefore the trivial spin structure
%% $(\begin{smallmatrix} \vec{0} \\ \vec{0} \end{smallmatrix})$
corresponds to choosing all $k_1^{{1}/{2}}$, $k_2^{{1}/{2}}$, and $\big[k(\mathbf{S}_1 \mathbf{S}_2^{-1})\big]^{{1}/{2}}$ negative, while $[k(\mathbf{S}_1 \mathbf{S}_2)]^{{1}/{2}}$, got by exchanging $\mathbf{u_2}$ with $\mathbf{v_2}$ in the above expression of  $\big[k(\mathbf{S}_1 \mathbf{S}_2^{-1})\big]^{{1}/{2}}$, turns out to be positive. This symmetry will be exploited later to define the non-separating complete degeneration. The GSO projection is implemented simply by summing over the four possible values of $\vec{\varsigma}$.

 In the presence of a constant background gauge field, the factors $F^{(0)}_{\rm gl}$
 and $\mathbf{F}_{\rm gl}^{(0)}$ in the measures of integration, in \eq{ghosts} and
 in \eq{ghostsf} respectively, get modified, since string coordinates with Neumann
 boundary conditions propagate in space-time and are sensitive to such backgrounds.
 The relevant modification to \eq{ghosts} was derived in Ref.~\cite{Magnea:2004ai}.
 One simply needs to substitute the factor $F^{(0)}_{\rm gl}$ in~\eq{aritwomeas} with
 the factor
 \beq
   F_{\rm gl}^{(\epsilon)} \, = \, \frac{ \big[ \det \left({\rm Im} \, \tau \right)
   \big]^{- \frac{d}{2} + 1}}{\det \left({\rm Im} \, \tau_\epsilon \right)}
   \frac{ {\rm e}^{- {\rm i} \pi \vec{\e} \cdot \tau \cdot \vec{\e} } \,
   {\prod_\alpha}' \prod_{n = 1}^\infty \left(1 - k_\alpha^n
   \right)^{- d + 2}}{{\prod_\alpha}' \prod_{n = 1}^\infty \left(1 - \ex{\, 2 \pi \ii
   \vec{\epsilon} \cdot \tau \cdot \vec{N}_\alpha} \, k_\alpha^n\right) \left(1 -
   \ex{- 2 \pi \ii \vec{\epsilon} \cdot \tau \cdot \vec{N}_\alpha} \, k_\alpha^n \right)}~.
 \label{eq:Fegl}
 \eeq
 % At $g$ loops, the vectors $\vec{\widetilde{\varsigma}}$ and $\vec{N}_\alpha$ have
 % $g$ components. The vector $\vec{\widetilde{\varsigma} } $ has components
 % $\widetilde{\varsigma }_\mu  = \varsigma_\mu  + 1$ (mod 2) where $\varsigma_\mu$
 % can take the value 0 or 1 defining the spin-structure along the $b$-cycles of the Riemann
 % surface;
 At $g$ loops, the vector $\vec{N}_\alpha$ %, on the other hand,
  has $g$ integer-valued entries: the $i$-th
 entry counts how many times the generator $S_i$ enters in the element of the Schottky
 group $T_\alpha$, with $S_i$ contributing $1$, while $S_i^{-1}$ contributes $-1$.

 The determinant of a new $(g-1)\times (g-1)$ matrix $\tau_\epsilon$, which reduces to
 $(\det \tau)$ when $\epsilon\to 0$, enters \eq{eq:Fegl}. Geometrically, the matrix
 $\tau_\epsilon$ can be seen as a `twisted' version of the period matrix, related to the
 periods along the $b$-cycle of the $(g - 1)$ regular differentials with the same monodromies
 as the string coordinates $X^\pm$ along the $a$-cycles. An explicit expression for
 these (Prym) differentials in the Schottky parametrization was derived
 in~\cite{Russo:2003tt,Magnea:2004ai}, and their periods were studied
 in~\cite{Magnea:2004ai,Russo:2007tc}. In the following, we will use the results
 in section~4.2 of~\cite{Magnea:2004ai}. Using the same techniques, developed and
 described in Refs.~\cite{Russo:2003tt,Magnea:2004ai,Russo:2007tc}, it is possible
 to generalize this construction to the Neveu-Schwarz spin structure of the RNS
 superstring.  The result takes the form
 %LORENZO% I have introduced \widetilde{\varsigma} here too.
 \beqa
   \mathbf{F}_{\rm gl}^{(\epsilon)} & = & {F}_{\rm gl}^{(\epsilon)}  \, \,
   {\prod_\alpha}' \prod_{n = 1}^\infty
   \left({1 -
   % \ex{\, \ii \pi \vec{\widetilde{\varsigma}} \cdot \vec{N}_\alpha} \,
   k_\alpha^{n - \frac{1}{2}}} \right)^{d - 2}
   \label{eq:Fegls}  \\
   && \times \, \left({1 -  \ex{ \, 2 \ii \pi % \left(\frac{\vec{\widetilde{\varsigma}}}{2} +
  % \,
  \vec{\epsilon} \cdot \tau %\right)
  \cdot \vec{N}_\alpha} \, k_\alpha^{\, n -
   \frac{1}{2}}} \right) \left({1 -  \ex{- 2 \ii \pi %\left(\frac{\vec{\widetilde{\varsigma}}}{2}
   % - \,
   \vec{\epsilon} \cdot \tau %\right)
   \cdot \vec{N}_\alpha} \, k_\alpha^{\, n -
   \frac{1}{2}}} \right) \,,  \nonumber
 \eeqa
 where, of course, now we have to use the super-Schottky group to calculate the
 multipliers $k_\alpha$ and all other objects present in~\eq{eq:Fegls}. This completes
 the list of the ingredients needed to compute the partition functions and study their
 field theory limits. It should perhaps be emphasized that, strictly speaking, the
 expressions just derived for the partition functions are not completely well defined,
 because of infrared divergences, which are particularly severe in the bosonic case due to
 the propagation of states with a negative squared mass. One could use the D-brane
 separation in space-time as a natural IR regulator, but in this work we are actually
 interested in the amplitude integrands, so we will not need to discuss the infrared
 regularization in detail. We can now move on to the analysis of the field theory limit.

 \section{The field theory limit}
 \label{fitl}

 The string world-sheet degenerates completely into a Feynman-like graph when
 all moduli approach one of the boundaries of moduli space. In particular, we focus
 on the degenerations corresponding to gauge theory diagrams, and analyze how the
 unique topology of a planar $g$-loop string configuration generates the different graph
 topologies that contribute to the $g$-loop field theory amplitude. Note that the
 dimension of moduli space ($3 g - 3 + n$ real parameters for open string amplitudes
 with genus $g \geq 2$ and $n$ vertex operator insertions), is equal to the number of
 propagators in a Feynman diagram with only cubic vertices and with the same number
 of loops and external states. Therefore, complete degenerations of  the string amplitude
 will yield Feynman-like graphs with cubic vertices only; graphs with quartic vertices,
 which also occur in realistic renormalizable field theories, will arise from partial
 degenerations, where some of the string moduli are still to be integrated over a finite
 region of moduli space.

 The central issue in analyzing the field theory limit of a string amplitude is the identification
 of the appropriate change of variables between the dimensionless string moduli and the
 dimensionful parameters characterizing the propagation of light degrees of freedom
 in Feynman graphs. Such a change of variables, which is expected to be different for
 each degeneration limit, must involve the string slope $\alpha'$, which is the only
 dimensionful parameter in the string theory. Once the change of variables pertaining
 to the selected graph topology has been performed, one may explicitly take the
 low-energy limit, $\alpha' \to 0$, and recover from each corner of moduli space the
 sum of all Feynman diagrams with the appropriate graph topology, contributing to the
 amplitude in the effective field theory.

 A detailed analysis of one-loop amplitudes~\cite{Frizzo:2000ez,Bern:1991aq,
 DiVecchia:1996uq} led to the identification of the appropriate change of variables at
 the one-loop level, where the moduli space for vacuum amplitudes is one-dimensional. In
 that case the gauge theory amplitudes were obtained by taking the Schottky multiplier $k
 \to 0$. This idea is intuitively appealing, since $k$ is related to the radius of the isometric
 circles of the Schottky generator $S$, and it is also strongly suggested by the sewing
 procedure used to construct multi-loop string amplitudes in the context of the operator
 formalism (see~\cite{DiVecchia:1988cy} and references therein). More precisely, one
 finds
 \beq
   \log k \, \equiv \, - \, \frac{T}{\alpha'} \, ,
 \label{naif}
 \eeq
 where $T$ is the sum of the Schwinger proper times associated to the propagators
 forming the loop in the field theory Feynman diagrams. This simple prescription (suitably
 generalized to include insertions of vertex operators for external string states) is
 sufficient to recover one-loop scattering amplitudes for low-lying string states, such as
 scalars (bosonic string tachyons) and gluons.

\begin{figure}
 \begin{center}
 \subfloat[]{ \includegraphics[scale=0.08]{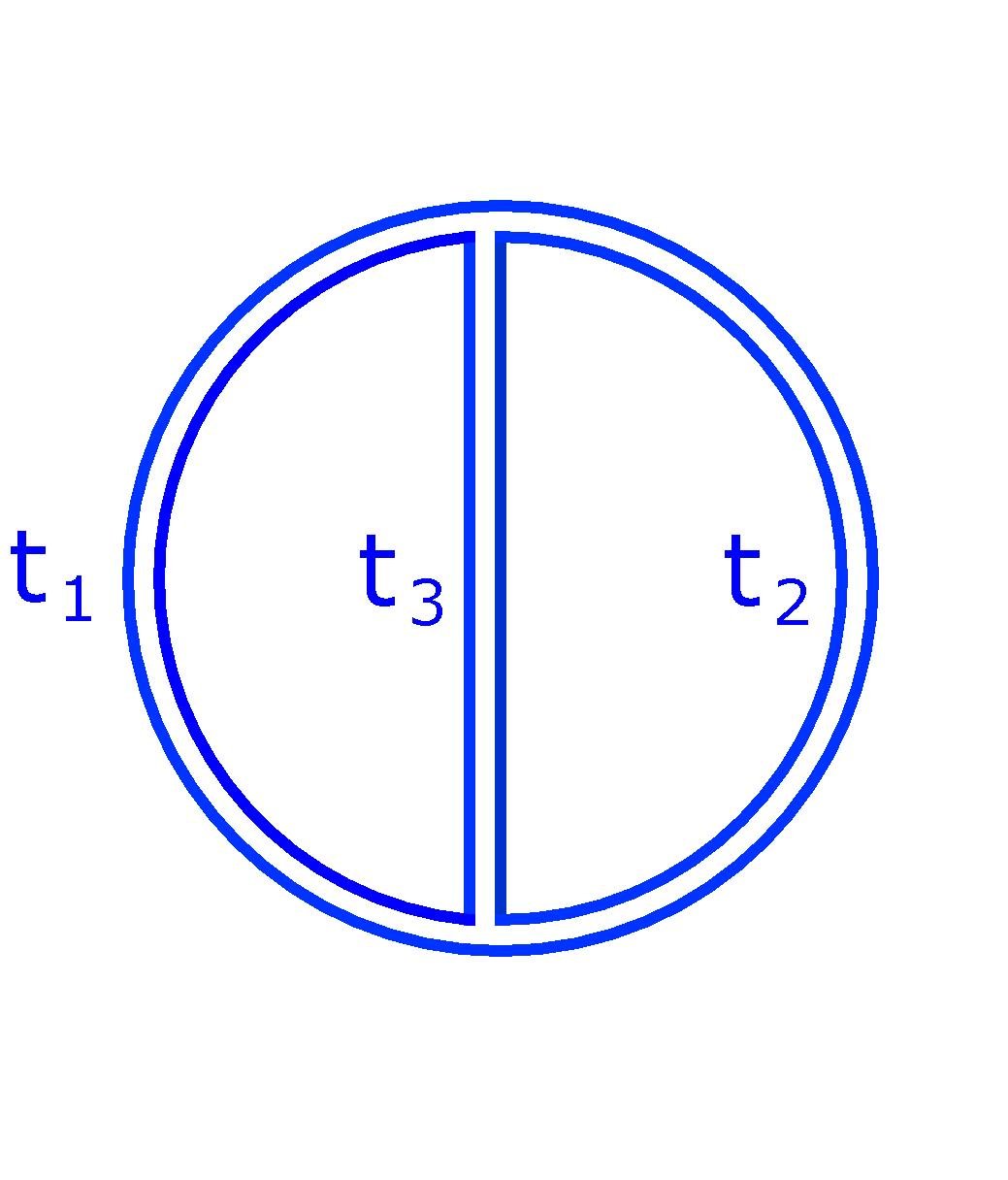}  \label{appletopology} }
 \subfloat[]{  \includegraphics[scale=0.08]{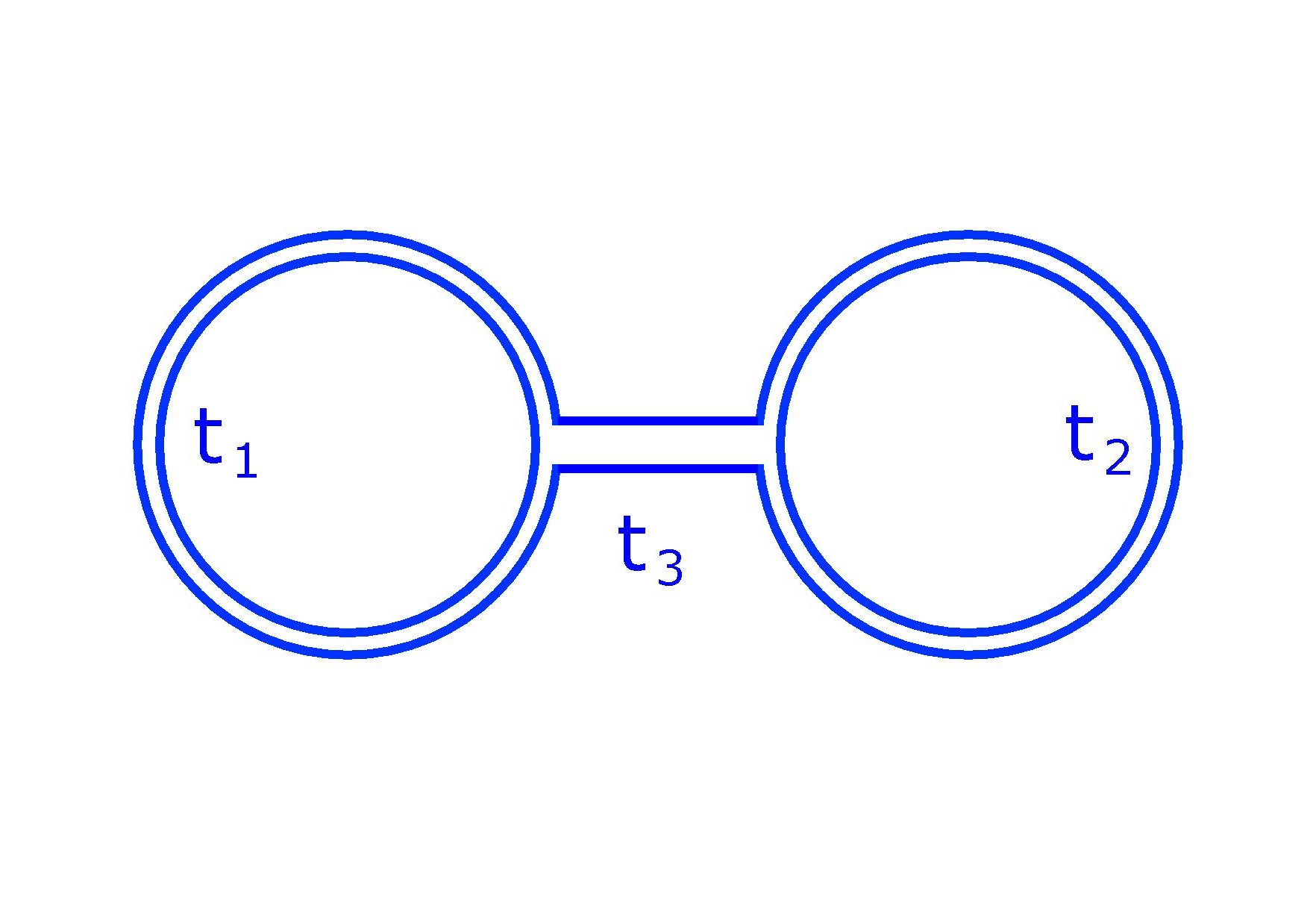}  \label{handlebartopology} }
 \subfloat[]{  \includegraphics[scale=0.08]{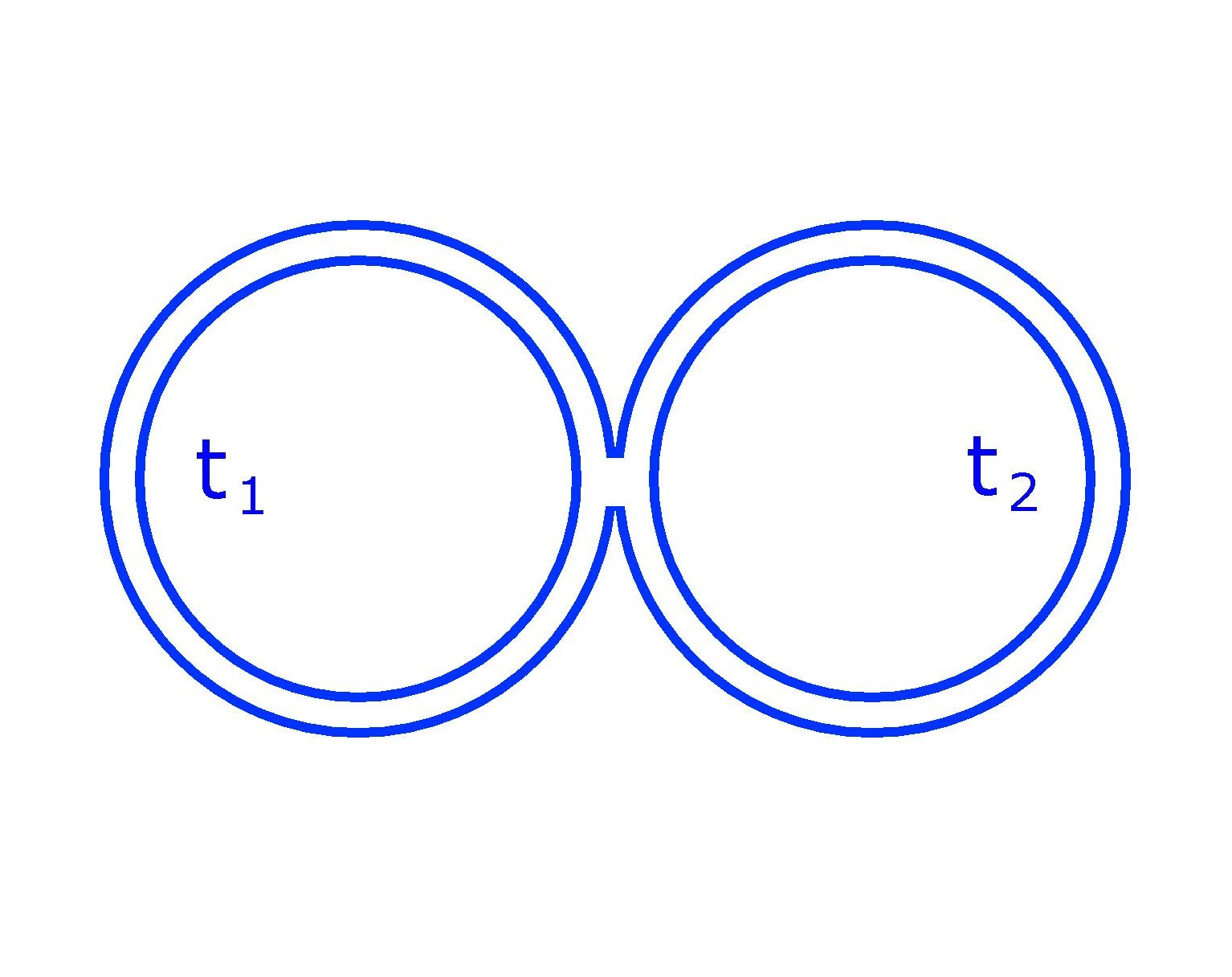}  \label{figeighttopology} }
 \caption{The three possible degeneration limits of the open string world-sheet for the
   two-loop vacuum amplitude. Individual propagators are labelled with their Schwinger
   proper-time variable $t_i$.}
 \label{topologies}
 \end{center}
 \end{figure}

 Beyond one loop, one must face the existence of different degeneration limits for
 vacuum amplitudes, corresponding to different graph topologies for vacuum Feynman
 diagrams. It is natural to expect that the field theory limit will still be driven by taking
 $k_i \to 0$ for $i = 1, \ldots, g$, while different configurations of  the fixed points will
 identify different graph topologies. This expectation was confirmed, to leading power
 in the multipliers, by the analysis of~\cite{Frizzo:1999zx,Magnea:2004ai,DiVecchia:1996kf}.
 To illustrate their results, consider the three distinct degeneration limits for two-loop
 vacuum string amplitudes, depicted in Fig.~\ref{topologies}, and naturally related to
 the three distinct two-loop Feynman graph topologies involving only cubic or quartic
 vertices.
 In these limits, a possible identification between moduli and Schwinger proper times
 $t_i$ for individual propagators is
 \beqa
   &&
   2 \pi \ii \tau_{11} \, \simeq \, \log k_1 \, = \, - \frac{t_1 + t_3}{\alpha'} \, , \qquad
   2 \pi \ii \tau_{22} \, \simeq \, \log k_2 \, = \, - \frac{t_2+t_3}{\alpha'} \, ,
   \nonumber \\
   &&
   \hspace{3cm} 2 \pi \ii \tau_{12} \, \simeq \, \log \eta \, = \, - \frac{t_3}{\alpha'} \,
 \label{fig1a}
 \eeqa
 for Fig.~\ref{appletopology}, and
 \beq
   2 \pi \ii \tau_{11} \, \simeq \, \log k_1 \, = \, - \frac{t_1}{\alpha'} \, , \qquad
   2 \pi \ii \tau_{22} \, \simeq \, \log k_2 \, = \, - \frac{t_2}{\alpha'} \, , \qquad
   \log \left( 1 - \eta \right) \, = \, - \frac{t_3}{\alpha'} \,
 \label{fig1b}
 \eeq
 for Fig.~\ref{handlebartopology}. In the case of the contact interaction of
 Fig.~\ref{figeighttopology}, no Schwinger proper time is associated to the modulus $\eta$.

 This prescription is enough to identify correctly the contributions of the
 (tachyonic) ground state for bosonic strings, and provides an explicit connections
 between two-loop string amplitudes and those of a $\phi^3$ field theory. One may
 easily verify that, as shown in Eqns.~(\ref{fig1a}) and~(\ref{fig1b}), the limit $\eta \to 0$
 corresponds to the `symmetric' topology shown in Fig.~\ref{appletopology}, while the
 limit $\eta \to 1$ yields the `handlebar' topology of Fig.~\ref{handlebartopology}, as discussed
 in Ref.~\cite{Frizzo:1999zx}. The first problem we wish to address now is how to
 generalize the prescriptions in Eqns.~(\ref{fig1a}) and~(\ref{fig1b}), so as to
 isolate the contributions of the massless degrees of freedom to the various available
 graph topologies at genus two and higher. Since massless states are not generally
 the lowest-energy states, corrections to Eqns.~(\ref{fig1a}) and~(\ref{fig1b}) which are
 subleading by powers of the multipliers may, and do, play a role.

 In the present case, the most interesting limit is the `symmetric' topology in
 Fig.~\ref{appletopology}, which we will consider in detail. Our first guiding principle
 is to connect the sums of all Schwinger proper times forming each loop to the
 multipliers of some element of the Schottky group. The second key point is that
 the dictionary between string and field theory quantities must respect the
 symmetry of the graph representing the degeneration of the Riemann surface,
 since this symmetry is a remainder of the modular invariance of the original string
 diagram. At two loops, for the symmetric topology, this is achieved by the following
 observation. In the Schottky parametrization, acting on the string boundary
 coordinate with the transformation $S_i$ corresponds to moving it around the
 $i$-th loop once, in a prescribed direction. Similarly, one easily verifies that acting
 with the transformation $S_1 S_2^{-1}$ corresponds to wrapping once around
 the outer boundary of the word sheet. In the symmetric topology, these three
 transformations must be interchangeable, and each one must be associated with
 the sum of two Schwinger proper times, since each field theory loop is composed
 of two propagators. These constraints are solved by imposing that
 \beq
   k (S_1) \, = \, p_1 \, p_3  \, , \qquad
   k (S_2) \, = \, p_2 \, p_3  \, , \qquad
   k (S_1 S_2^{-1}) \, = \, p_1 \, p_2 \, ,
 \label{pi}
 \eeq
 where each $p_i$ is then associated with a single field theory propagator, and to
 the corresponding proper time variable, according to
 \beq
   \log p_i \, = \, - \, \frac{t_i}{\alpha'} \, .
 \label{Schwin}
 \eeq
 Remarkably, the change of variables from the set $\{k_1, k_2, \eta\}$
 to the set $\{ p_i \}$ is simple, and one finds
 \beq
   \eta \, = \, \frac{(1 + p_1)(1 + p_2) p_3}{(1 + p_3) (1 + p_1 p_ 2 p_ 3)} \, ,
 \label{etapi}
 \eeq
 which allows us to expand the partition functions in~\eq{zeps} in a Laurent series
 in powers of $p_i$ without encountering singularities. At leading power in $p_i$, the
 assignment in \eq{Schwin} coincides with that in~\eq{fig1a}. Notice however that,
 already at next-to-leading power in any of the $p_i$, the two choices differ.

 In the supersymmetric theory, the approach is similar, but there are also important
 differences. First of all, the various factors in the partition function must be expanded
 in powers of $\sqrt{p_i}$, instead of $p_i$, as can be seen for example in \eq{ghostsf}.
 %%%
 %%%
 This means we need to extend equation \eq{pi} to account for the different signs $k_i^{{1}/{2}}$. According to the convention explained after~\eqref{ghostsf} we specify
 \begin{align*}
 k_1^{{1}/{2}} & =  - \sqrt{p_1} \sqrt{p_3} &
 k_2^{{1}/{2}} & =  - \sqrt{p_2} \sqrt{p_3} &
 k_3^{{1}/{2}} = \big[ k (S_1 S_2^{-1})\big]^{{1}/{2}} & = - \sqrt{p_1} \sqrt{p_2}
 \end{align*}
where all of the $\sqrt{p_i}$'s are positive. 

 %%%
 %%%
 More importantly, the standard procedure to extract the field theory limit can be applied
 only after having performed the Berezin integrations over supermoduli. As noted
 in~\cite{Witten:2013cia}, it then becomes essential to choose correctly the bosonic
 moduli that are kept fixed when the Berezin integration is performed. In the present
 case, one may see that the correct bosonic variable is the quartic invariant $y$ in
 \eq{eta}, and not the closely related one $\eta$. The reason for this is that the bosonic
 parameter describing the gluing of two punctured annuli that forms our two-loop
 surface is in fact $y$~\cite{Witten:2013cia}. In the Schottky approach this can be
 seen by repeating, in the supersymmetric case, the sewing procedure discussed
 in Ref.~\cite{Cristofano:1988cf}.
 In order to extract the field theory limit in the superstring case, we must thus begin
 by writing the measure of integration in terms of $y$, before performing the integration
 over the Grassmann parameters. In fact, we expect this prescription to be valid for all
 values of $y$ that yield a well-defined Riemann surface, and not just around the
 degeneration point $y = 0$. These arguments, which arise from an explicit construction
 of the two-loop Riemann surface, find a confirmation when the field theory
 limit is explicitly computed: as we will see in the next section, where we consider
 the non-separating degeneration $y \to 1$, by using $\eta$ instead of $y$ one gets
 a result which is inconsistent with basic properties of the low-energy effective theory.
 By sewing, one constructs an explicit expression
 for the Schottky generators $\hat{\mathbf S}_1 $, $\hat{\mathbf S}_2$ of the
 disk with two holes, as obtained by gluing two
 annuli with punctures at ${\mathbf z}_1=(1,\theta_1)$ and ${\mathbf
 z}_2=(1,\theta_2)$, by means of a propagator
 of `length' $\log y$. As in the bosonic case~\cite{Cristofano:1988cf}, one finds
 \beq
   \hat{\mathbf S}_1 \, = \, {\mathbf P}(y)^{-1} \, {\mathbf V}_1^{-1} \,
   \mathbf{P}(k_1) \mathbf{V}_1 \mathbf{P}(y) \, , \qquad
   \hat{\mathbf S}_2 \, = \, \mathbf{\Gamma} \, {\mathbf V}_2^{-1} \,
   \mathbf{P}(k_2) \mathbf{V}_2  \, \mathbf{\Gamma} \, ,
 \label{eq:genCNP}
 \eeq
 where
 $\mathbf{P}(k_i)$ are the generators of the two annuli and $\mathbf{P}(y)$ is
 the propagator which performs the gluing; moreover $\mathbf{V}_i$ define local
 super-coordinates around the two punctures, and the matrix
 \beq
   {\mathbf \Gamma} \, = \, \left(
   \begin{array}{ccc}
     0 & 1 & 0 \\
     1 & 0 & 0 \\
     0 & 0 & 1
   \end{array}\right) \, .
 \label{eq:Gamma}
 \eeq
 exchanges the origin with the point at infinity.

 One can choose the simplest local coordinates, using the supertranslations
 \beq
   {\mathbf V}_i \, = \, \left(
   \begin{array}{ccc}
     1 & 1 & - \theta_i \\
     0 & 1 & 0 \\
     0 & \theta_i & 1
   \end{array}\right) \, , \qquad
 \label{supertr}
 \eeq
 which send the origin to the punctures at ${\mathbf z}_i$, and one can use the simplest
 propagator
 \beq
   {\mathbf P}(y) \, = \, \left(
   \begin{array}{ccc}
      - \sqrt{y} & 0 & 0 \\
      0 & \frac{-1}{ \sqrt{y}} & 0 \\
      0 & 0 & 1
   \end{array}\right) \, .
 \label{eq:tadgl}
 \eeq
 In this framework, the Schottky generators $\hat{\mathbf S}_1 $, $\hat{\mathbf
 S}_2$ are related by a similarity transformations to the ${\mathbf S}_1 $,
 ${\mathbf S}_2$ we will choose in Eq.~\eqref{eq:gfv1}. The gluing parameter $y$
 in \eq{eq:genCNP} is exactly the supersymmetric version of the quartic invariant $y$
 introduced in~\eq{eta} and is related to the other quartic invariant $\eta$ as
 \beq
   \eta \, = \, \big( \mathbf{u}_1, \mathbf{u}_2, \mathbf{v}_1, \mathbf{v}_2
   \big) \, = \,
   1 - y + \sqrt{\eta} \, \, \Theta_{\mathbf{v}_1 \mathbf{u}_1 \mathbf{u}_2}
   \Theta_{\mathbf{v}_1 \mathbf{u}_1 \mathbf{v}_2} \, .
 \label{etay}
 \eeq
 We can now summarize the prescription for studying the complete degeneration
 leading to the graph in Fig.~\ref{appletopology}: first we have to expand each factor
 in the measure in a Laurent series in powers of each of the $p_i$'s (or $\sqrt{p_i}$
 in the NS case); then we have to pick the term that cancels all tachyonic poles, so
 as to have a measure proportional to $d p_i/p_i = - d t_i/\alpha'$: indeed, the operator
 formalism of Ref.~\cite{DiVecchia:1988cy} suggests that in general the propagation
 of states at the $n$-th mass level of the string should correspond to the term of
 order $n - 2$ in this Laurent series. Finally, in the presence of a background gauge
 field, we have to rewrite $\epsilon$ in terms of the physical field $B$, by using~\eq{epsi}.
 At this stage, the field theory integrands for the Feynman diagrams are recovered by
 taking the limit $\alpha' \to 0$, with $B$ and the $t_i$'s kept fixed.

 By taking the field theory limit as described, string theory yields directly the sum of
 all Feynman diagrams with the chosen graph topology. For example, both in the
 bosonic and in the NS cases, each of the three degenerations in Fig.~\ref{topologies}
 yields a sum of diagrams involving gluons, ghosts, and adjoint scalars. Furthermore,
 various particles can be charged or neutral with respect to the selected background
 field, and charged vector particles can be polarized in the `magnetized' $(x_1, x_2)$
 plane, or in one of the $d - 2$ `unmagnetized' directions. One can, however, perform
 a more detailed analysis. By tracing back the origin of the various factors contributing
 to the string theory partition function, it is actually possible to disentangle the
 contributions to each graph topology of each individual particle state, thus extracting
 directly from string theory the integrands of individual Feynman diagrams. As a
 byproduct, this leads to the identification of the gauge in which the field theory
 calculation should be performed to reproduce these results.

 The idea is simple. Each factor in the partition functions in~\eq{zeps} gives a
 specific contribution, corresponding to the propagation of certain space-time
 fields and polarizations, which can be identified by recognizing how the factor
 originates from the path integral over world-sheet fields. First of all, it is natural to
 identify the factor $F_{\rm scal}$ ($\mathbf{F}_{\rm scal}$ in the supersymmetric
 case) with the contribution of space-time adjoint scalars, since this factor arises
 from functional integration over string coordinates with Dirichlet boundary conditions.
 Thus, if we extract a factor of $p_i$ ($\sqrt{p_i}$, or a factor quadratic in Grassmannian
 supermoduli, for the NS case) from this term, we may conclude that the propagator
 associated with the proper time $t_i$ describes a scalar in the corresponding Feynman
 diagram. Similarly, the factor arising from the functional integral over the world-sheet
 ghosts $b$ and $c$ must correspond to the space-time propagation of Faddeev-Popov
 ghosts in the low-energy gauge theory: this identification is supported by comparing
 the world-sheet and space-time BRST transformations, as done for example in
 section~4.3 of Ref.~\cite{Polchinski:1998rq}; we conclude that the expansion of $F_{\rm
 gh}$ ($\mathbf{F}_{\rm gh}$) will select space-time ghost fields in the corresponding
 field theory propagators. Finally, the factor $F_{\rm gl}$ ($\mathbf{F}_{\rm gl}$),
 arising from functional integration over string coordinates with Neumann boundary
 conditions, yields the gluon contribution to the partition function. In fact, we can
 distinguish between Feynman diagrams coming from different gluon polarizations:
 to find the two degrees of freedom parallel to the magnetized ($x_1$ $x_2$) plane,
 we extract $p_i$ from the $\epsilon$-dependent denominator of Eq.~(\ref{eq:Fegl}),
 and to get the $(d - 2)$ perpendicular degrees of freedom we extract $p_i$ from
 the $\epsilon$-independent terms in the numerator. These give different contributions
 even if the gluon itself is uncharged with respect to the background field, because of
 its covariant derivative coupling to charged particles.

 While we have described in detail the choice of variables appropriate to the degeneration
 in Fig.~\ref{appletopology}, our approach is completely general and can be extended to
 other degeneration limits and to higher genus as well. The guiding principles are always
 the fact that multipliers must be associated with sums of proper time variables for all
 propagators comprising the loop, and the residual modular symmetries of the diagram
 must be properly taken into account. These ideas can, for example, be used to analyze
 the other complete degeneration points at two loops (such as for instance the `handlebar'
 limit of Fig.~\ref{handlebartopology}), or partial degenerations capturing contributions of
 diagrams with four-point vertices, as in Fig.~\ref{figeighttopology}. We will discuss in
 detail the variables appropriate to these cases, and the comparison with the corresponding
 field theory diagrams, in a forthcoming paper.

 \section{Extracting Feynman diagrams: an example}
 \label{exam}

 To substantiate our arguments in the previous section, we now proceed to
 extract from the string partition function, in the non-separating degeneration
 limit depicted in Fig.~\ref{appletopology}, the contribution corresponding to a chosen
 individual Feynman diagram in the low-energy theory. As an example, we select the
 contribution of charged Faddeev-Popov ghosts circulating in the two external
 propagators parametrized by $t_1$ and $t_2$, with a gluon state neutral under
 the background field $B$ propagating in the middle line parametrized by $t_3$.
 This choice is simple enough to allow for a short derivation and a direct comparison
 between the bosonic string and the superstring formalisms, and it already contains
 information about the gauge choice implicit in the string calculation. We postpone the
 presentation of our detailed analysis of the other diagrammatic contributions to
 the effective action to a future publication.

 As discussed in the previous section, for the symmetric topology in Fig.~\ref{appletopology}
 we need to use the variables $\{p_i\}$ defined in~\eq{pi}. In the bosonic case, it is
 straightforward to rewrite the zero-mode part of the ghost measure $F_{\rm gh}$ in
 \eq{ghosts} in terms of these variables. One finds
 \beq
   \frac{dk_1}{k_1^2} \, \frac{dk_2}{k_2^2} \,
   \frac{d\eta}{(1 - \eta)^2} (1 - k_1)^2 (1 - k_2)^2 \, = \,
   \frac{dp_1}{p_1^2} \, \frac{dp_2}{p_2^2} \, \frac{dp_3}{p_3^2} \,
   (1 - p_2 p_3)(1 - p_1 p_3)(1 - p_1 p_2) \, .
 \label{eq:ZkZp}
 \eeq
 One immediately notices the complete symmetry of this expression under permutations
 of the $p_i$'s, which is a pre-requisite for the consistency of the field theory limit for the
 symmetric topology. One also notes the absence of terms linear in only one of the $p_i$'s,
 which corresponds to the absence of diagrams with a single ghost propagator in the
 gauge theory. To pick our chosen diagram, we must compensate the quadratic poles
 in $p_1$ and $p_2$ by taking the term linear in $p_1 p_2$ from $F_{\rm gh}$. This
 is completely given in \eq{eq:ZkZp}, as one easily sees that the infinite product in
 the first line of \eq{ghosts} contributes terms which are at least quadratic in the
 $p_i$'s. In order to complete the calculation, we must next select the term linear in
 $p_3$, but without any powers of $p_1$ and $p_2$, from the gluon factor $F_{\rm
 gl}^{(\epsilon)}$. In order to do this, we need the approximate expressions for the only
 two objects that remain non-trivial when we set $p_1$ and $p_2$ to zero: the determinant
 of the period matrix $\det \tau$, and its twisted version, $\det \tau_\epsilon$. We can extract
 these results from~\cite{Magnea:2004ai,Russo:2007tc}. Up to corrections vanishing
 as powers of $p_1$ or $p_2$, the elements of the period matrix are
 \beq
   2 \pi i \tau_{1 1} \, = \,  \log (p_1 p_3) \, , \qquad
   2 \pi i \tau_{2 2} \, = \,  \log (p_2 p_3) \, , \qquad
   2 \pi i \tau_{1 2} \, = \, 2 \pi i \tau_{21} \, = \, \log \eta \, .
 \label{eq:tau}
 \eeq
 Furthermore, the expression for the determinant of the twisted period matrix, at
 leading power in $p_1$ and $p_2$ and including the linear term in $p_3$, is given in the
 limit $\epsilon_2 \to -\epsilon \, , \, \epsilon_1 \to \epsilon$, as
 \beqa
   {\rm Im} (\tau_\epsilon) & = &
   \frac{1}{4 \pi \sin(\pi \epsilon)} \, \Bigg[ \frac{1}{\epsilon} \,
   \left( p_1^{\epsilon/2} - p_1^{- \epsilon/2} \right) \left( p_2^{\epsilon/2} -
   p_2^{- \epsilon/2} \right) + \, \Big(2 \gamma_E + \log p_3
   + \psi(-\epsilon) + \psi(\epsilon)\Big)
   \nonumber \\
   && \times \, \left(\left( p_1^{\epsilon/2} p_2^{\epsilon/2} - p_1^{- \epsilon/2}
   p_2^{- \epsilon/2} \right)-\epsilon p_3 \left(p_1^{\epsilon/2} p_2^{-\epsilon/2} + p_1^{-\epsilon/2}p_2^{\epsilon/2} \right)\right) \Bigg]~, \label{eq:taue}
 \eeqa
 %LORENZO% inserted - \psi(1) below
 where $\psi(z)=\Gamma'(z)/\Gamma(z)$ is the digamma function and $\gamma_E = - \psi(1)$
 is the Euler-Mascheroni constant. Both expressions must be written in terms of $p_i$, and
 to extract our chosen diagram we must further expand in powers of $p_3$ and pick the
 term linear in $p_3$. The leading order in $p_3$ of~\eq{eq:taue} is given by Eq.~(4.7)
 of~\cite{Russo:2007tc}; it is possible to generalise that derivation to include also the
 subleading term which is proportional to $p_3$. However there is an alternative, simpler
 approach and we will come back to this point in a subsequent paper. Assembling the various
 factors, and including the overall normalization $C_g (\vec{\epsilon})$, it is then straightforward
 to take the $\alpha' \to 0$ limit, which yields the expression for our chosen diagram,
 which we label by $D_{\rm gh}^{\rm sym} (B)$. We find
 \beq
   D_{\rm gh}^{\rm sym} (B) \, = \, K \, \frac{g^2}{(4 \pi)^d} \, \int_0^\infty
   \left[\prod_{i = 1}^3 d t_i \right]
   \frac{t_3}{\Delta_0^{d/2} \, \Delta_B}
   \Bigg\{ \frac{d - 2}{2}
   + \frac{\Delta_0}{\Delta_B} \, \cosh \big(B  (t_1 - t_2) \big)
   \Bigg\} \, ,
 \label{ex}
 \eeq
 where $g$ is the gauge theory coupling, $K$ is a normalization factor to be discussed below,
 and we introduced the notations
 \beqa
    \Delta_B & \equiv & \frac{1}{B^2}  \Big[ \sinh(B t_1) \sinh(B t_2) \, + \,
    t_3 B \sinh \big[ B (t_1 + t_2) \big] \Big] \, , \nonumber \\
    \Delta_0 & \equiv & \lim_{B \to 0} \Delta_B \, = \, t_1 t_2 + t_1 t _3  + t_2 t_ 3 \,  .
 \label{deltas}
 \eeqa
These are related to the period matrix and to the twisted period matrix, after making the
substitutions in~\eq{epsi} and \eq{Schwin}, via
\beq
 \Delta_B  \, =  \, \lim_{\alpha' \to 0} \, \left[ \left( 2 \pi \alpha' \right)^2 \,
 \det \left({\rm Im} \tau_\epsilon \right) \right] \, , \qquad
 \Delta_0 \, = \, \lim_{\alpha' \to 0} \, \left[ \left( 2 \pi \alpha' \right)^2 \, {\rm det}
 \left( {\rm Im} \tau \right) \right] \, .
\label{deltodet}
\eeq
 In~\eq{ex}, one easily identifies the first term in braces as related to the $d-2$ gluon
 polarizations along `neutral' directions, while the remaining term comes from gluons
 polarized in the `magnetized' $(x_1 x_2)$ plane.

 In quantum field theory, the precise value, and in fact the very existence, of diagrams
 like the one just discussed, depends of course on the gauge choice. For the particular
 diagram we are considering, it turns out that the structure of the integrand in~\eq{ex}
 follows from the standard choice of the Feynman gauge, see~\cite{Pawlowski:2008xh} (the
 diagram we considered corresponds to the terms with an explicit ${\cal F}^2$ in the integrand
 ${\cal I}_{\rm ghost}$ in~\cite{Pawlowski:2008xh}; the remaining contributions are related to
 the diagram with a neutral ghost and a charged gluon). Even for this simple diagram,
 however, in order to recover the correct normalization we find that the gauge choice
 must be modified. In order to identify the gauge, we have performed the calculation of
 the effective action in the background field $B$, using the background field method, as
 described in~\cite{Abbott:1980hw}, but with the non-linear covariant gauge-fixing choice
 suggested by~\cite{Bern:1991an}, which evolves out of the Gervais-Neveu gauge first
 identified in Ref.~\cite{Gervais:1972tr}. While leaving a detailed analysis of the gauge
 choice on the field theory side to future work, we note here that the correct choice is
 to pick the gauge fixing Lagrangian
 \beq
   {\cal L}_{\rm GF} \, = \, - \, {\rm Tr} \left[ {\cal G}^2 \left( {\cal A},
   {\cal Q} \right) \right] \, ,
 \label{lgf}
 \eeq
 where ${\cal A}$ is the background field, ${\cal Q}$ the quantum field, and
 \beq
   {\cal G} \left( {\cal A}, {\cal Q} \right) \, = \, D_\mu^{\cal A} {\cal Q}^\mu + \frac{\rm i}{2}
   \gamma g \left\{ {\cal Q},{\cal Q} \right\} \, ,
 \label{gnbf}
 \eeq
 with $D_\mu^{\cal A}$ the covariant derivative with respect to the background field.
 If one computes, for example, the diagram $D_{\rm gh}^{\rm sym} (B)$ in this gauge,
 one finds precisely the structure of \eq{ex}, with an overall normalization $K = 1 +
 \gamma^2$. We have checked that the string theory results reproduce all gauge theory
 diagrams, as for the ghost diagram, with the choice $\gamma^2 = 1$ (note that the sign
 of $\gamma$ is immaterial, since $\gamma$ appears only quadratically in all diagrams).

 Let us now outline the analysis of the same example from the superstring
 point of view. As discussed in the previous section, the Berezin integration should
 be defined by keeping constant the bosonic variable $y$. Instructed by the bosonic
 case, we know that the complete degeneration depicted in Fig.~\ref{appletopology} is
 described in terms of the variables $p_i$, where however the definition in \eq{pi} now
 involves the multipliers of super-Schottky transformations. We need thus to write the
 supersymmetric analogue of~\eq{etapi}. For for simplicity, we fix our global super-projective
 coordinates by choosing $\mathbf{u_1} = (0, 0)$, $\mathbf{v_1} = (\infty, 0)$, $\mathbf{u_2}
 = (u, \theta)$, and $\mathbf{v_2} = (1, \phi)$. This implies
 \beq
   \eta \, = \, u \, , \
 \Theta_{\mathbf{v}_1 \mathbf{u}_1 \mathbf{u}_2}
   \, = \,  \theta/\sqrt{u} \, , \qquad  \Theta_{\mathbf{v}_1 \mathbf{u}_1 \mathbf{v}_2}
   \, = \, \phi \, .
 \label{eq:gfv1}
 \eeq
 %% If we use $k_3$ to indicate the multiplier of the super-Schottky transformation
 If we use $k_3^{{1}/{2}}$ to denote the smallest (in absolute value) eigenvalue
 of
 $\mathbf{S}_1 \mathbf{S}_2^{-1}$ (so $k_3$ is its multiplier), we can express $y$ as
 \beq
   y \, = \, \frac{(1-k_1)(1-k_2) \, + \, \theta \, \phi \, \Big[
   (1 - \ex{\ii \pi \varsigma_1} k_1^{{1}/{2}}) (1 - \ex{\ii \pi \varsigma_1} k_2^{{1}/{2}})  (1 + \ex{\ii \pi (\varsigma_1+\varsigma_2)} k_1^{{1}/{2}} k_2^{{1}/{2}}) \Big]}{1 + k_1 k_2 - k_1^{{1}/{2}}k_2^{{1}/{2}}( k_3^{{1}/{2}} + k_3^{-{1}/{2}})} \, .
 \label{etapis}
 \eeq
 With our choice of coordinates, the prefactor in the supersymmetric measure
 of integration given in square brackets in~\eq{aritwomeasf} takes a particularly
 simple form\footnote{Notice that the change of variables from $k_1$, $k_2$ and $y$ to $p_1$, $p_2$ and $p_3$ does not introduce in the measure any factor dependent on the Grassmann variables $\theta$ and $\phi$ as $d \log y/d k_3$ is independent of $\theta$ and $\phi$.},
 and can be written as
 \beq
   \frac{d k_1}{k_1^{3/2}}  \, \frac{d k_2}{k_2^{3/2}} \, \, d \log y  \, d \theta \, d \phi
   \, = \, \frac{d p_1}{p_1^{3/2}} \, \frac{d p_2}{p_2^{3/2}} \, \frac{d p_3}{p_3}
   \, d \theta \, d \phi \, \, \frac{1- p_1 p_2}{(1 + p_3) (1 + p_1 p_2 p_3)} \, .
 \label{aritwomeasf2}
 \eeq
 Contrary to the bosonic result,~\eq{eq:ZkZp}, this expression is not symmetric
 under permutations of the $p_i$, and it does not become symmetric even after
 including the zero-mode contribution of the ghost sectors, given by the first factor in
 $\mathbf{F}_{\rm gh}$ in~\eq{ghostsf}. Fortunately, we can restore the symmetry, and
  the one-to-one correspondence between string and field theory integrands, by
 exploiting the freedom to rescale Grassmann variables in the Berezin integral.
 We do so by defining
 \beq
   \hat \theta_{i j} \, = \, c_{i j} \, \Theta_{\mathbf{v}_i \mathbf{u}_i
   \mathbf{u}_j} \, , \qquad
   \hat \phi_{i j} \, = \, c_{i j} \, \Theta_{\mathbf{v}_i \mathbf{u}_i
   \mathbf{v}_j} \, ,
 \label{eq:newtheta}
 \eeq
 where
 %LORENZO% rewritten eqn below, perhaps in a `nicer' way. Below, also noted
 %that \varsigma_3 = \varsigma_1 + varsigma_2 (from Stefano).
 \beq
   c_{1 2} \, = \,  \left[ \left(1 +  \ex{\ii \pi \varsigma_3} \sqrt{p_1}\sqrt{p_2} \right)
   \left(1 - \ex{\ii \pi \varsigma_1} \sqrt{p_1}\sqrt{p_3} \right) \left(1 - \ex{\ii \pi
   \varsigma_2} \sqrt{p_2}\sqrt{p_3} \right) \right]^{-1/2} \, ,
 \label{c12}
 \eeq
 with $c_{23}$ and $c_{31}$ obtained by permuting the indices $(123)$. As before,
 $\varsigma_3=\varsigma_1+\varsigma_2$, $\mathbf{u}_3$ and $\mathbf{v}_3$ label
 respectively the spin structure and the fixed points of the transformation $\mathbf{S}_1
 \mathbf{S}_2^{-1}$. In terms of these new Grassmann variables,~\eq{aritwomeasf2},
 multiplied times the zero-mode contribution of the ghost sectors in~\eq{ghostsf}, reads
 \beq
   \prod_{i = 1}^3 \left[ \frac{d p_i}{p_i^{3/2}} \,
   \frac{1 + \ex{\ii \pi \varsigma_i}k_i^{{1}/{2}}
   }{\sqrt{1 + p_i}} \right]  d \hat\theta_{12} \, d \hat\phi_{12} \,
   \frac{1}{\sqrt{1 + p_1 p_2 p_3}} \,  ,
 \label{eq:ZkZps}
 \eeq
 which is the supersymmetric analogue of~\eq{eq:ZkZp}. One can check that
 $d \hat\theta_{12} d \hat\phi_{12} \, = \, d \hat\theta_{23} d \hat\phi_{23} \, = \,
 d \hat\theta_{31} d \hat\phi_{31}$, so that~\eq{eq:ZkZps} is fully symmetric under
 permutations of the super-Schottky transformations $\mathbf{S}_i$ as expected.

 We can now easily derive the result in~\eq{ex} from the superstring partition
 function $\mathbf{Z}(B)$ in~\eq{zeps}. We first isolate the term proportional to
 $\sqrt{p_1 p_2}$ in the ghost contribution~\eqref{eq:ZkZps}. As
 before, this corresponds to selecting the contribution of space-time ghosts in
 the external loop of the diagram in Fig.~\ref{appletopology}. Next we must expand
 $\mathbf{F}_{\rm gl}^{(\epsilon)}$ to first order in $\sqrt{p_3}$ and to zeroth
 order in $\sqrt{p_1}$ and $\sqrt{p_2}$. As in the bosonic case, terms of this type
 can arise only from the expansion of the super period matrix, both in the twisted
 and in the untwisted version. As before, the first case corresponds, on the field
 theory side, to contributions due to the propagation, in the middle propagator of
 Fig.~\ref{appletopology}, of a gluon neutral with respect to the background field
 $B$, and with a polarization parallel to the magnetised plane, while the second
 case corresponds to perpendicular polarizations. It is straightforward to
 supersymmetrize the standard and the twisted period matrix and repeat the steps outlined
 in~\cite{Magnea:2004ai,Russo:2007tc} to calculate the superperiods at zeroth
 order in $p_1$ and $p_2$. Indeed, at this order, the result for the untwisted period matrix in terms of $\eta$
 is identical to the bosonic one, provided super fixed points and supersymmetric
 differences $\braket{\mathbf{z}_1}{\mathbf{z}_2}$ are used in place of bosonic
 ones. On the other hand, at this order the twisted superperiod matrix has a term linear in $\theta \phi$, when written in terms of $\eta$. Our final ingredient is the expression for $\eta$ in the supersymmetric
 case: by using~\eq{eq:gfv1} and~\eq{etapis} we find
 \beq
   \eta \, = \, \frac{p_3}{1 + p_3} (1 + \theta \phi) + {\cal O}(\sqrt{p_1},\sqrt{p_2})
   \, = \, {p_3} \left(1 + \sqrt{p_3} \, \hat\theta_{12} \, \hat\phi_{12}\right) + {\cal
   O} (\sqrt{p_1}, \sqrt{p_2}, p_3^2) \, .
 \label{eq:up3s}
 \end{equation}
 This has to be compared with~\eq{etapi} in the bosonic case, after expansion in
 powers of $p_3$, which yields $\eta = p_3 (1 - p_3) + {\cal O}(p_1, p_2, p_3^3)$. The integrands have both analytic and logarithmic (e.g.~$\eta^\epsilon$) dependence on $\eta$.
 The term coming from the logarithmic dependence of $\mathbf{F}_{\rm gl}^{(\epsilon)}$ on $\eta$ which is linear  in  $\hat\theta_{12} \hat\phi_{12}$ has the same structure as the first-order term coming from the logarithmic dependence in the expansion
 of ${F}_{\rm gl}^{(\epsilon)}$, except for the fact that it is proportional to $\sqrt{p_3}$
 instead of $p_3$. It also turns out that the linear term in $ \theta \phi$ in the twisted superperiod matrix has the same form as the linear term in $\eta$ in the twisted period matrix, so when we write $\theta \phi = \sqrt{p_3} \hat\theta_{12} \hat\phi_{12} + {\cal O}(\sqrt{p_1}, \sqrt{p_2}, p_3)$ we see that the untwisted and twisted superperiod matrices are almost identical except the linear terms in $p_3$ are instead proportional to $\sqrt{p_3} \hat\theta_{12} \hat\phi_{12}$. This, however, is exactly what we need in order to compensate the
 tachyonic pole in the superstring measure in~\eqref{eq:ZkZps}, so that finally we
 reproduce exactly the field theory result in~\eq{ex}.

 Let us conclude this section by analysing what happens if we change our local
 coordinates on supermoduli space and choose, for instance, to fix the bosonic
 modulus $\eta$ (instead of $y$) in the Berezin integration. It is straightforward to
 check that this corresponds to the choice of Ref.~\cite{D'Hoker:2002gw}\footnote{Actually
   Ref.~\cite{D'Hoker:2002gw} focuses on closed superstring theory; their results,
   however, are written in a holomorphically factorised way and spin structure by
   spin structure. It is thus easy to extract the open string measure corresponding to
   our $[d \mathbf{m}]^0_2$.}, at least at leading order in the $\tau_{ij} \to \ii \infty$
 limit, for the two-loop vacuum amplitude at $\epsilon = 0$. With this choice, it is not
 possible to trace a one-to-one correspondence between each Feynman diagram
 and individual contributions to the string amplitudes. However, even without
 aiming at separating different diagrams, we can still ask whether we can interpret
 from a field theory point of view the {\em leading} contribution to the symmetric graph
 topology in the $B \to 0$ limit. We would expect to find a correspondence with a diagram
 similar to the one discussed in this paper, where ghost lines have been substituted
 by NS tachyon propagators, while the third propagator is still a gluon. Now, the main
 feature  of the result obtained by keeping $\eta$ fixed is that the integrand of each
 NS spin structure in the complete degeneration limit is proportional to $\Delta_0^{- d/2}$.
 Any sensible interaction term with two scalar particles and a gluon must however involve
 a derivative, which inevitably leads to a numerator linear in the appropriate Schwinger
 parameter in the diagram integrand. In field theory, it thus appears impossible to
 generate this dependence on $\Delta_0$, which is typical of non-derivative couplings.
 This is to be contrasted with the integrand obtained by using $y$ in the Berezin
 integration, which is proportional to $t_3/\Delta_0^{d/2 + 1}$, as one finds in the
 $\epsilon\to 0$ limit of~\eq{ex}, and as expected from the coupling structure of the
 low-energy gauge theory.

 \section{Conclusions}
 \label{pers}

 In this paper we have shown how to parametrize a Riemann surface describing
 a degenerating multi-loop string amplitude, so as to make explicit contact with the
 corresponding Feynman diagrams of the low-energy quantum field theory. We have
 been working in the RNS formalism, and we found the Schottky parametrization
 of (super) Riemann surfaces to be especially suited to study the correspondence
 between string and field theory amplitudes. We have proposed an essentially
 algorithmic procedure that allows us to start from the full expression for an
 open string amplitude, as an integral over (super) moduli space, and derive from
 it the corresponding gauge theory amplitude on a diagram-by-diagram
 basis. In the present paper we focused on issues concerning the parametrization
 of the string world sheet, and on the comparison between the bosonic string
 setting and the NS sector of the superstring. To this end, we used, as an example
 of our technique, the study of string effective actions in a constant background
 gauge field, and, in the process, we also presented some new results concerning
 the multi-loop expression for this effective action for the NS spin structure of the
 superstring. To illustrate our results, we have presented a relatively simple example,
 deriving explicitly from the two-loop string partition function a specific two-loop
 Feynman diagram involving a loop of space-time ghosts and a gluon propagator.
 We leave to forthcoming papers a detailed analysis of specific 4D field
 theories, and the discussion of several issues relevant to the field theory description
 emerging from the string amplitudes, such as the regularization of IR divergences,
 the precise gauge fixing procedure, and the most convenient approach to
 renormalisation. In this concluding section we would like to make some general
 remarks about what can be learnt from the study of multi-loop (super)string
 amplitudes in the degeneration limit.

 A first interesting point is that, in the low energy limit, we are able to obtain from
 string amplitudes a set of field theory quantities, such as the Schwinger-parameter
 integrands of individual Feynman diagrams, which contain off-shell information, allowing
 for example to infer the Feynman rules of the low-energy theory. We note that this relation
 works with the specific set of coordinates we used in the degeneration limit, and with a
 specific choice of gauge on the field theory side. Perhaps not surprisingly, the gauge
 we identified is a natural generalization of the non-linear gauge found in the study
 of tree-level string amplitudes in Ref.~\cite{Gervais:1972tr}, as first suggested in
 Ref.~\cite{Bern:1991an}. Interestingly, it does not seem possible to modify in a simple
 way the approach on either side of the correspondence. For instance, as discussed
 at the end of the previous section, a different choice for Berezin integration yields,
 in the complete degeneration limit, results that are difficult to interpret from the field
 theory point of view. Similarly, there is no obvious way to change the choice of gauge
 one finds on the field theory side, at least within the framework of covariant string
 quantization. We recall that, on the other hand, strings quantized on the light
 cone~\cite{Goddard:1973qh,Thorn:2002fj} result in quantum-field theory amplitudes
 computed in a physical light-cone gauge.

 Another important feature of our result is that it is very flexible, and works in setups
 with a different number of Neumann and Dirichlet directions, and different brane
 configurations. Here we considered the case where the low energy theory describing
 massless open string states is a Yang-Mills theory in $D = 26$ ($D = 10$ super
 Yang-Mills theory in the superstring case), reduced to $d$ space-time dimensions.
 We note that actually the number $s$ of scalar states (and thus the number of
 space-time supersymmetries in the case of RNS string theory) present in the
 low-energy theory can be easily tuned by introducing orbifold projections to reduce
 their number below the one dictated by the critical dimension. In a similar vein, we
 expect that our approach can be extended to higher-genus Riemann surfaces, and
 to scattering amplitudes with massless external states. It is reasonable to expect that similar techniques could be used to calculate closed string amplitudes, yielding expressions for graviton scattering amplitudes in a variety of gravity theories.

 Finally, even in the simplest case, where no orbifold projection is considered, it is
 interesting to study explicitly the twisted string partition function discussed in this paper
 for different values of $d$. For instance, while $d = 4$ is obviously relevant for making
 contact with four-dimensional gauge theories, the choice $d = 1$ describes the
 dynamics of D$0$-branes. In this setting, one should analytically continue the twist parameter
 $\epsilon \to \ii \epsilon $ to describe the relative velocities between different D$0$-branes; this
 case was studied in detail in the context of the BFFS model in Ref.~\cite{Banks:1996vh}.
 At the one-loop level we know exact string results~\cite{Bachas:1995kx,Billo:1997eg}
 that interpolate between the open string/M(atrix) limit and the closed string/gravity
 limit. At two loops, the two limits were studied in detail in~\cite{Okawa:1998pz}, but
 no full string derivation is known. After including the contribution of the Ramond spin
 structures, our approach should simplify the comparison between the full two-loop
 string partition function and the formulae describing the interaction of three moving
 D$0$-branes in an appropriate limit; this should provide stringent checks on whether
 all the subtleties of RNS multi-loop amplitudes have been properly resolved. We
 plan to address this range of interesting open issues in our future work.

 \vspace{1cm}

 \noindent {\large \bf Acknowledgements.}

 \vspace{2mm}

 \noindent We would like to thank C.~Bachas and P.~Di Vecchia for enlightening discussions.
 This work was supported by STFC (Grant ST/J000469/1, `String theory, gauge
 theory \& duality') and by MIUR (Italy) under contracts 2006020509$\_$004 and
 2010YJ2NYW$\_$006.  L. M. and S. Sc. are very grateful to CERN and to QMUL for
 hospitality during the completion of this work; L. M. also thanks NIKHEF (Amsterdam)
 for hospitality and support. R. R. wishes to thank Universit\'e Pierre et Marie Curie (Paris)
 and Universit\`a di Torino for hospitality, and the `Institut Lagrange de Paris' for
 support during the completion of this work. S. P. also thanks Universit\`{a} di Torino for
 hospitality.

%%%%%%%%%%%%%%%%%%%%%%%%%%%%%%%%%%%%%%%%%%%%%

\bibliographystyle{JHEP}
\bibliography{rs}

%%%%%%%%%%%%%%%%%%%%%%%%%%%%%%%%%%%%%%%%%%%%%

\end{document}